\documentclass[pra,twocolumn]{revtex4}
\usepackage{epsfig}
\usepackage{graphicx}
\usepackage{amsmath}
\usepackage{natbib}
\usepackage{color}

\newcommand{\n}{\nonumber}
\newcommand{\bn}{\begin{eqnarray}}
\newcommand{\en}{\end{eqnarray}}
\newcommand{\eml}{\end{multline}}
\newcommand{\bml}{\begin{multline}}
\newcommand{\h}{\hspace}
\newcommand{\vs}{\vspace}

\begin{document}

\title {Measurement and Significance of Wilson Loops in Synthetic Gauges Fields}

 \author{Kunal K. Das}
 \affiliation{
Department of Physical Sciences, Kutztown University of Pennsylvania, Kutztown, Pennsylvania 19530, USA\\
Department of Physics and Astronomy, State University of New York, Stony Brook, New York 11794-3800, USA}

\date{\today }
\begin{abstract}
We study Wilson loops as a necessary tool for unambiguous identification of non-Abelian synthetic gauge fields, with attention to certain crucial but often overlooked features, such as the requirement of at least three distinct loops.  We devise a method to determine the complete Wilson loop matrix from the time evolved amplitudes of the internal atomic states of laser-coupled ultracold atoms that does not require lattice confinement.  The analysis is done in the context of a new cyclic model that can realize both Abelian and non-Abelian structures within a single configuration with continuous variation possible between U(1) and U(2) gauge groups by varying the  detuning of the laser fields.
\end{abstract}
%

\maketitle

\section{Introduction}

In a seminal paper \cite{berry}, Berry noted that the geometric phase acquired by a quantum state during adiabatic evolution displays features of an U(1) gauge field, an observation swiftly generalized by Wilczek and Zee to non-Abelian counterparts \cite{Wilczek-Zee-PRL-non-abelian}. In recent years, that connection has found utility in creating synthetic gauge structures for access to a broad range of fundamental physics phenomena in systems of ultracold atoms \cite{RMP-Dalibard,Review-Goldman,Dum-Olshanii-PRL,Bergmann-PRA-STIRAP-geometric, Fleischhauer-PRL-non-abelian-gauge,Clark-zitterbewegung, Juzeliunas-double-refraction,Galitski-spin-orbit,Jaksch-Zoller,Spielman-magnetic, Spielman-Zitterbewegung, Spielman-non-Abelian-monopole}.
These new developments have also brought forth a surprising diversity of opinions about how to identify non-Abelian gauge fields. Different studies over the years have used a plethora of criteria, including presence of degeneracy \cite{Bergmann-PRA-STIRAP-geometric}, non-vanishing commutators of the vector potential \cite{Fleischhauer-PRL-non-abelian-gauge,Clark-zitterbewegung,Juzeliunas-double-refraction} and numerical value of a Wilson loop \cite{Goldman-Lewenstein-Wilson-loop}. Even contradictory viewpoints were manifest in separate cold atom studies of the same phenomenon \cite{Ketterle-butterfly}.  There has been progress on resolving some of these differences using arguments based on field and loop variables \cite{Review-Goldman}, however mostly in the context of lattices, and as will be shown here, they are incomplete.

The primary goal of this paper is therefore to present a broadly applicable criterion to distinguish  truly non-Abelian synthetic gauge fields that works even without a lattice, and to provide a procedure to implement that criterion in experiments with cold atoms. We will demonstrate by direct simulation, its utility in identifying and resolving ambiguities of other criteria in use. Our analysis is facilitated by our secondary goal of introducing a novel cyclic scheme that can create both Abelian and non-Abelian structures within the same configuration.

We will show that Wilson loops \cite{Wilson-PRD,Zee-PRA}, when properly evaluated and interpreted can provide the necessary criterion.  Important in lattice gauge theories as a nonlocal gauge invariant observable, Wilson loops have been considered only for lattices in the context of ultracold atoms \cite{Bloch-wilson-lines,Goldman-Lewenstein-Wilson-loop,Review-Goldman}. While their value in examining gauge commutativity have been noted \cite{Goldman-Lewenstein-Wilson-loop,Review-Goldman}, certain essential factors were overlooked that obscured their full utility. That has been compounded by the fact that in studies of synthetic gauge fields, Wilson loops continue to be associated with physical loops in lattices, and otherwise remain an abstract concept, not easy to measure in experiments. We remedy that here by providing a simple method to measure the Wilson loop as well as the complete associated matrix, that significantly does not require a lattice and is therefore applicable to a broader range of experiments on artificial gauge fields.

\section{Wilson Loops}

Consider a time-dependent Hamiltonian $H(t)$ which for all relevant time evolutions contains a subspace $\{\Phi_i(t)|i\in[1,2,\cdots,N]\}$  of instantaneous dark states \cite{Shore-book} that satisfy $H(t)\Phi_i(t)=0$ for the static Hamiltonian at every instant. When $H(t)$  varies slowly on the scale of the inverse energy gap separating the dark states from adjacent states, the description can be consistently confined to the subspace of dark states, and the restricted state vector represented by them, $\Psi_i(t)=W_{ij}(t) \Phi_j(t)$. Its index signifies the initial state $\Psi_i(0) = \Phi_i(0)$. Insertion into the Schr\"odinger equation leads to coupled equations for the amplitudes,
\bn \label{evolution-equations} \dot{W}_{ij} =i \vec{A}_{ik} W_{kj}\cdot\dot{\vec{\mu}}, \h{5mm} A_{ik}=i\langle \Phi_i|\nabla|\Phi_k\rangle.\en
where components of the vector $\vec{\mu}$ are system parameters. The matrix $\vec{A}$, in general, transforms like a non-Abelian vector potential $\vec{A}\rightarrow U\vec{A}U^\dagger-i(dU)U^\dagger$ under a local unitary transformation of the dark state basis. In the special case, when $N=1$, $\vec{A}$ transforms like a $U(1)$ Abelian gauge potential. Formal integration leads to a path-ordered (${\cal P}$) integral for the evolution matrix,
\bn  W={\cal P} e^{i\int d\vec{\mu}\cdot \vec{A} },\h{4mm}W_{\circ}={\cal P} e^{i\oint d\vec{\mu} \cdot \vec{A} },\h{4mm} {\cal W}= {\rm tr}[W_{\circ}]\label{Wilson}, \en
in the parameter space. In the adiabatic limit, the evolution is unitary. The line integral $W$ depends on the choice of gauge. But, its value over a closed loop, $W_{\circ}$, when traced, ${\cal W}$, is a gauge invariant quantity known as the Wilson loop \cite{Wilson-PRD}. In order to differentiate the former from its trace, we will refer to $W_{\circ}$ as the Wilson loop matrix. Two points are worth noting, first, the gauge structures reside in the space of parameters that typically are not spatial coordinates, second, in the adiabatic limit assumed, time only serves to mark progress, not its rate, along the parametric path.

\section{Hamiltonian and States}
 For our simulations we introduce an effective four state system described by a time-dependent Schr\"odinger equation $i\hbar\partial_t \Psi= H \Psi$ with the interaction Hamiltonian
\bn\label{Hamiltonian} H=
\frac{\hbar}{2}\left(
\begin{array}{cccc}
 0 & e^{i \varphi_1} p & 0 & e^{-i \varphi_4} p \\
 e^{-i \varphi_1} p & \delta & e^{i \varphi_2} q & 0 \\
 0 & e^{-i \varphi_2} q & 0 & e^{i \varphi_3} q \\
 e^{i \varphi_4} p & 0 & e^{-i \varphi_3} q & \mp \delta
\end{array}
\right)\en
with state vector specified by the complex amplitudes of the bare states $\Psi(t) =(C_a,C_b,C_c,C_d)$. We will set $\hbar=m=1$, where $m$ is the mass of the specific atom species used, and we assume energy, length and time units, $\epsilon_0=\Omega_0$, $\tau_0=\Omega_0^{-1}$ and $l_0=(\Omega_0)^{-1/2}$, based on some characteristic frequency $\Omega_0$ at the scale of the Rabi oscillations for the relevant atomic levels.
If the phases satisfy $ \sum_i \varphi_i=2\pi n$ for integer $n$, this Hamiltonian has the convenient property, that depending on the sign of the detuning in $H_{44}$ there are one or two zero eigenvalues and corresponding dark states
\bn\label{non-Abelian-eigen}
H_{44}=-{\textstyle\frac{1}{2}} \delta:\h{5mm}&&\{0 , 0 , \pm{\textstyle\frac{1}{2}} \sqrt{\delta^2+2 \left(p^2+q^2\right)}\},\\
H_{44}=+{\textstyle\frac{1}{2}} \delta:\h{5mm}&& \{0, {\textstyle\frac{1}{2}} \delta,{\textstyle\frac{1}{4}} (\delta\pm\sqrt{\delta^2+8 \left(p^2+q^2\right)}) \}.\n
 \en
This allows realization of both non-Abelian and Abelian synthetic gauge structures within the same configuration.

The matrix elements can be parameterized by
\bn\label{parametrization}
\varphi_1=\alpha,\  \ \varphi_2=\beta,\h{.87cm} &&\varphi_1+\varphi_4=-\varphi_2-\varphi_3=\gamma\n\\
\Omega=\sqrt{\delta^2+2(p^2+q^2)}&& \delta=\Omega\sin(\phi)\n\\
p={\textstyle\frac{1}{\sqrt{2}}}\Omega\sin(\theta)\cos(\phi)&&
q={\textstyle\frac{1}{\sqrt{2}}}\Omega\cos(\theta)\cos(\phi),\en
we assume all values to be non-negative real numbers. When $H_{44}=-{\textstyle\frac{1}{2}} \delta$, an orthonormal set of basis vectors spanning the sub-space of the two dark states are
\bn\label{non-Abelian-dark}  \Phi_1&=&\left(e^{i\alpha} \cos(\theta),0,-e^{-i \beta}\sin(\theta),0\right)\n\\
\Phi_2&=&\left(e^{i \alpha} \sin(\phi) \sin(\theta),-{\textstyle \frac{1}{\sqrt{2}}}\cos(\phi),\right.\n\\&&
\left.\ \ \ e^{-i\beta} \sin(\phi) \cos(\theta),e^{i\gamma}{\textstyle \frac{1}{\sqrt{2}}}\cos(\phi)\right).\en
When $H_{44}=+{\textstyle\frac{1}{2}} \delta$, the degeneracy is lifted, and $\Phi_1$ retains the same form and is the sole dark state, while $\Phi_2=\left(0,
-{\textstyle \frac{1}{\sqrt{2}}},0,{\textstyle \frac{1}{\sqrt{2}}}e^{i \gamma}\right)$ corresponds to the eigenvalue $\delta/2$.  As the  detunings are varied continuously $(\delta,-\delta)\rightarrow (0,0)\rightarrow (\delta,\delta)$, the dark states decouple at $(0,0)$, and the gauge structure changes, $U(2)\rightarrow U(1)\times U(1)$.

\section{Gauge Potential and Field} For a single dark state, $\Phi_1$, only the relative phase of the two non-vanishing components matters, so we set $\beta=0$. As recognized in numerous studies \cite{RMP-Dalibard,Review-Goldman}, the entity $ A_{\alpha}=i \langle \Phi_1|\partial_{\alpha}|\Phi_1\rangle=-\cos^2\theta$ (note $A_\theta=0$) acts like a vector potential in the Schr\"odinger equation, transforming like an Abelian $U(1)$ gauge potential when the state is multiplied by a co-ordinate dependent phase factor. The corresponding field is $F_{\alpha\theta}=\partial_\alpha A_\theta-\partial_\theta A_\alpha= -\sin(2\theta)$.

But, with two dark states, the degeneracy leads to a $2\times 2$ matrix vector potential with components,
\bn  A_\theta&=&-\sin\phi\ \sigma_y, \h{1cm}
  A_\phi=0\n\\
  A_\alpha&=&-{\textstyle \frac{1}{2}}\sin\phi \sin(2\theta)\ \sigma_x -\cos^2\theta\ \sigma_\uparrow
-\sin^2\phi \sin^2\theta\ \sigma_\downarrow \n\\
   A_\beta&=&-{\textstyle \frac{1}{2}}\sin\phi \sin(2\theta)\ \sigma_x+\sin^2\theta\ \sigma_\uparrow
+\sin^2\phi \cos^2\theta\ \sigma_\downarrow\n\\
   A_\gamma&=&-{\textstyle \frac{1}{2}} \cos^2\phi\ \sigma_\downarrow,\en
in the space of parameters $\mu \in \{\theta,\phi,\alpha,\beta, \gamma\}$. They transform as components of a $U(2)=U(1)\times SU(2)$ gauge potential and represented here in terms of the generators, $I_2$ the identity and $\sigma_{i=x,y,z}$ the Pauli spin matrices, along with projection operators $\sigma_{\uparrow(\downarrow)}=\frac{1}{2}(I_2\pm \sigma_z)$. The  corresponding gauge field components are given by
\bn\label{Field}  F_{\mu\nu}=\partial_\mu A_\nu-\partial_\nu A_\mu- i[ A_\mu, A_\nu].\en
With five parameters, there are $^5C_2=10$ distinct non-trivial combinations. For ease of distinguishing their origins, we list separately all the \emph{non-vanishing} contributions from the curl and from the commutator,
\bn \label{curls}\partial_\theta A_\phi-\partial_\phi A_\theta&=&-\cos\phi\ \sigma_y\\
\partial_\phi A_\alpha-\partial_\alpha A_\phi&=&-{\textstyle\frac{1}{2}}\cos\phi \sin(2\theta)\sigma_x-\sin(2\phi)\sin^2\theta\   \sigma_\downarrow\n\\
\partial_\phi A_\beta-\partial_\beta A_\phi&=&-{\textstyle\frac{1}{2}}\cos\phi \sin(2\theta)\sigma_x+\sin(2\phi)\cos^2\theta\    \sigma_\downarrow\n\\
\partial_\phi A_\gamma-\partial_\gamma A_\phi&=& {\textstyle\frac{1}{2}}\sin(2\phi)\ \sigma_\downarrow\n\\
 \partial_\theta A_\alpha-\partial_\alpha A_\theta&=& \partial_\theta A_\beta-\partial_\beta A_\theta\n\\ \h{-1.2cm}&&\h{-1.2cm}=-\sin\phi\cos(2\theta)\sigma_x
+ \sin(2\theta) [\sigma_\uparrow-\sin^2\phi\ \sigma_\downarrow],\n\en
\bn\label{commutators} -i[ A_\theta, A_\alpha]
&=&\sin\phi \left(\cos^2\theta-\sin^2\phi \sin^2\theta\right)\sigma_x\n\\&&  -\sin^2\phi \sin(2\theta)\sigma_z\n\\
-i[ A_\theta, A_\beta]&=&\sin\phi \left(\cos^2\theta \sin^2\phi-\sin^2\theta\right)\sigma_x \n\\&&-\sin^2\phi \sin(2\theta)\sigma_z\n\\
-i [ A_\theta, A_\gamma]&=&-{\textstyle\frac{1}{2}}\sin\phi\cos^2\phi\ \sigma_x \n\\
-i [ A_\alpha, A_\beta]&=&{\textstyle\frac{1}{2}}\sin\phi\cos^2\phi \sin(2\theta)\ \sigma_y \n\\
-i [ A_\alpha, A_\gamma]&=&-i [ A_\beta, A_\gamma]=-{\textstyle\frac{i}{2}}[ A_\alpha, A_\beta].\en
Here, commutators involving $A_\phi$ vanish as well as the components of the curl with only phase degrees of freedom. A vector field can be constructed by allowing the parameters to have spatial variation \cite{RMP-Dalibard}, here we consider temporal variation instead.

\begin{figure}[t]
\centering
\includegraphics[width=\columnwidth]{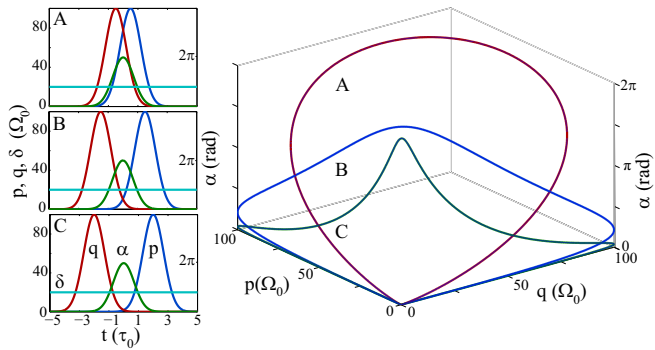}
\caption{(Color online) Time evolution implemented by Gaussian pulses for the parameters, $p,q,\alpha$, and constant $\delta$. Three loops, labeled A, B, C are shown corresponding to three different delays $\tau=0.5,1.5,2\ \tau_0$ between the pulses. }
\label{Figure-2}
\end{figure}

\section{Path and Evolution} The full set of parameters provide substantial flexibility, a restricted case will suffice here where we set $\beta=\gamma=0$, so the only non-vanishing vector potential components are $A_\theta$ and $A_\alpha$. In our simulations, the detuning will be held constant at $\delta=20$ and the remaining parameters varied in time with Gaussian profiles,
\bn h(t)=h_0e^{-(t-\tau_h)^2/\sigma_h^2}\h{1cm} h\in\{p,q,\alpha\}.\en
All pulse widths are set to be $\sigma_h=1$, relative delays to satisfy $\tau_p-\tau_\alpha=\tau_\alpha-\tau_q=\tau$, and amplitudes $p_0=q_0=100$ and $\alpha_0=2\pi$. The initial state will be $ \Psi(0)=|a\rangle $ so that $C_a(0)=1$, the time evolution of the state can be represented by the complex amplitudes $\{C_i(t)|i=a,b,c,d\}$. When $\tau>0$, the $q$ coupling precedes $p$ and vice versa for $\tau<0$, the analogs of counterintuitive and intuitive sequence in a lambda scheme \cite{Shore-book}, but here $\delta$ and $\alpha$ bridge the $p$ and $q$ pulses. In what follows, we consider evolution of the parameters in loops starting from and returning to zero, different loops created by changing $\tau$, as shown in Fig.~\ref{Figure-2}.

\begin{figure}[t]
\centering
\includegraphics[width=\columnwidth]{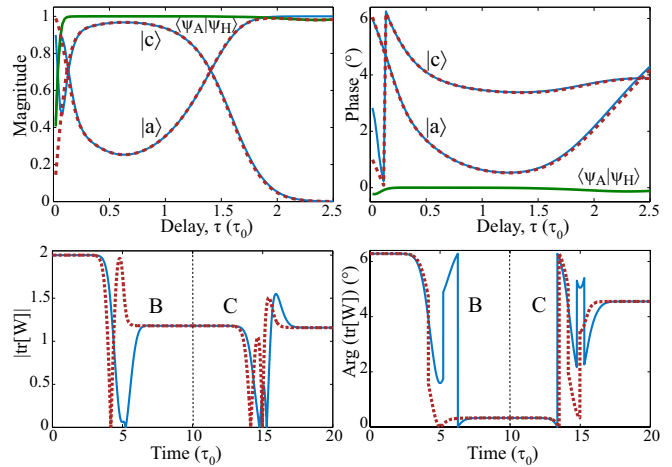}
\caption{(Color online)  \emph{Upper panels}: Comparison of the magnitude and the phase of the \emph{final} states due to adiabatic (A) and Hamiltonian (H) evolution, as the loop is varied by changing the delay $\tau$ between pulses, shows excellent agreement for their overlap $\langle \Psi_A|\Psi_H\rangle$ (solid green  line, almost constant) and for the amplitudes of the bare states $|a\rangle$ and $|c\rangle$ (A $\rightarrow$  solid blue line; H $\rightarrow$ dashed red line). \emph{Lower panels}: The two traces correspond to Wilson line integrals, without (dotted red line) the $U(2)$ gauge transformation in Eq.~(\ref{gauge-transformation}) and with it (solid blue line); they coincide at the end of each cycle, $t=10$ and $20\ \tau_0$, illustrating gauge invariance of the Wilson loop (B and C refer to the respective loops shown in Fig.~\ref{Figure-2}).}
\label{Figure-3}\vs{-.5cm}
\end{figure}

Adibaticity is confirmed for a range of such loops in Fig.~\ref{Figure-3} by comparing adiabatic (A) evolution via Eq.~(\ref{evolution-equations}) with evolution by the full Hamiltonian (H) in Eq.~(\ref{Hamiltonian}). The final state vectors at $t=T$ after a circuit, obtained by the two methods show excellent agreement, their overlap $\langle\Psi_H(T)|\Psi_A(T)\rangle$ having magnitude $\sim 1$ and phase $\sim 0$. Likewise, the amplitudes $C_a(T), C_{c}(T)$ of the levels with significant population at completion, are indistinguishable between the two ways of evolving, except when the pulses almost coincide or hardly overlap.

Gauge invariance of the Wilson loop  is illustrated by comparing the evolution of ${\rm tr[W]}$ with and without a $U(2)$ gauge transformation by
$U^\dagger=((1, 0),(0,e^{i\zeta(t)}))$ with $ \zeta(t)=\frac{1}{2}\alpha(t)$, whereby,
\bn &&A_\theta d\theta+A_\alpha d\alpha\rightarrow U A_\theta U^\dagger d\theta+U A_\alpha U^\dagger d\alpha +A_\zeta d\zeta\n\\
&&A_\zeta=-i(d_\zeta U) U^\dagger=-\sigma_\downarrow.\label{gauge-transformation}\en
Figure~\ref{Figure-3} shows that the evolution of ${\rm tr[W]}$ is affected, but at the end of each closed loop ${\cal W}={\rm tr[W_\circ]}$ stays invariant, both in magnitude and phase.

\section{Necessity for Wilson Loops} For an arbitrary set of loops $(A,B,C,\cdots)$, starting and ending at the same point in parameter space, if the net Wilson loop ${\cal W}(ABC\cdots)$ depends on their order, the gauge structure is non-Abelian but otherwise effectively Abelian. The utility of Wilson loops to accurately identify non-Abelian synthetic gauge fields can be appreciated by first highlighting the limitations of other criteria in vogue:

(I) Presence of a degenerate subspace ($N>1$) has been used as non-Abelian signature in some studies \cite{Bergmann-PRA-STIRAP-geometric,Ketterle-butterfly}. While that is a necessary condition, it is not sufficient, as is easily demonstrated by a counterexample. We set $\alpha=0$ in our model, that still leaves two coupled degenerate states, but only one non-vanishing component $A_\theta$, so all commutators vanish in Eq.~(\ref{curls}). Crucially, since $A_\theta \propto \sigma_y$ a single generator, $[A_\theta(t_1),A_\theta(t_2)]=0$ for any two points labeled by times $t_1$ and $t_2$. This allows the Wilson loop matrix to be evaluated analytically,
\bn W_\circ(\Lambda)=I_2\cos(\Lambda)+i\sigma_y\sin(\Lambda), \h{3mm} \Lambda=-{\textstyle \oint} d\theta\sin(\phi).\en
It is easily seen that $[W_\circ(\Lambda_A),W_\circ(\Lambda_B)]=0$, hence also their trace, for any two arbitrary closed loops. The field also has one component $F_{\theta\phi}\propto \sigma_y$ and thus commutes in all gauges due to its covariance described in (III) below.

(II) Non-vanishing commutators of the vector potential components are often used to label non-Abelian fields \cite{RMP-Dalibard,Fleischhauer-PRL-non-abelian-gauge,Clark-zitterbewegung,Juzeliunas-double-refraction}. Although, it is another necessary condition, definitive conclusions cannot be based on such commutators, since they are not invariant or even covariant under a gauge transformation, $A\rightarrow UAU^\dagger-i(dU)U^\dagger$. So, while all commutators could vanish in one gauge, they may not in another. This is easily illustrated with the counterexample in (I) where all commutators vanish. But, now apply the local gauge transformation we used before to illustrate gauge invariance of Wilson loops, $U^\dagger=((1, 0),(0,e^{i\zeta(t)}))$.  Then in the new gauge, the vector potential has two components $A_\theta=-e^{i\zeta}\sin(\phi)\sigma_y$ and $A_\zeta=-\sigma_\downarrow$ which clearly do not commute $[A_\theta,A_\zeta]\neq 0$.

(III) The field strength has been suggested as an alternate to remedy the gauge dependence of the latter $\vec{A}$. But, the field is not gauge invariant except when Abelian, rather it is gauge covariant, $F=dA-iA^2\rightarrow UFU^\dagger$ and that is only due to mutual cancelations of terms arising from the curl and the commutator,
\bn dA\rightarrow UdAU^\dagger-UAdU^\dagger+dUAU^\dagger+idUdU^\dagger\n\\
-iA^2\rightarrow -iUA^2U^\dagger+UAdU^\dagger-dUAU^\dagger-idUdU^\dagger,\en
where we used notation of exterior calculus. There would still remain the practical challenge of how to identify the contribution of the commutator in any measurement of the field. Besides, unlike the Wilson loop, the field is a local variable, so measuring commutators of its components at different points is non-trivial at best.

\begin{figure}[t]
\centering
\includegraphics[width=\columnwidth]{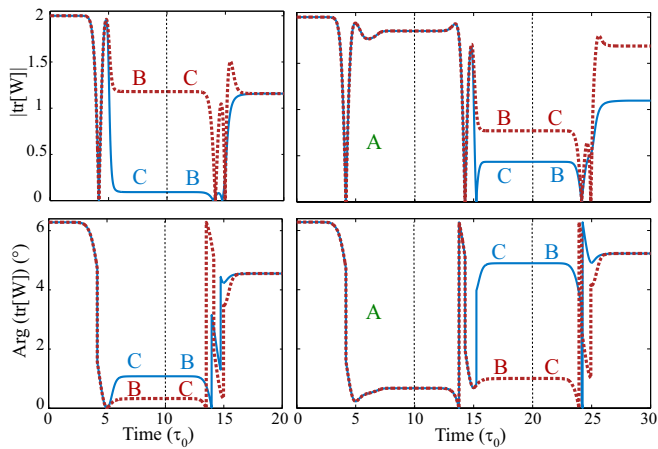}
\caption{(Color online) The evolution of the trace of the Wilson line integral is plotted, $|{\rm tr}[W]|$ (\emph{upper panels}) and ${\rm Arg}({\rm tr}[W])$ (\emph{lower}), over two loops (\emph{left}) and three loops (\emph{right}), the end of each loop marked by vertical dashed line (A,B and C refer to the loops shown in Fig.~\ref{Figure-2}). The \emph{left} panels show that with only two loops, the net Wilson loop is unchanged on switching their order $B,C$  (dotted red line) $\rightarrow C,B$ (solid blue line).  The \emph{right} panels show that with three loops, when the last two are switched $A,B,C$  (dotted red line) $ \rightarrow A,C,B$ (solid blue line), the net Wilson loops have the same phase but markedly different magnitudes.}
\label{Figure-4}
\vspace{-.5cm}
\end{figure}

\section{Application of Wilson Loops} In a few recent works on synthetic gauge fields, Wilson loops have indeed been discussed in the specific context of lattices \cite{Review-Goldman,Goldman-Lewenstein-Wilson-loop}. We now show that the conclusions were incomplete and demonstrate a fully gauge-invariant way to use Wilson loops to identify non-Abelian gauge structures, even with no lattice.

(IV) For $N$-fold degeneracy, it was proposed \cite{Goldman-Lewenstein-Wilson-loop} that the magnitude of the Wilson loop $|{\cal W}|\neq N$ signifies non-Abelian. In the case discussed in (II) with $A_\alpha=0$, it is effectively Abelian, yet $|{\cal W}|=2\cos(\Lambda) \leq N=2$. This not being a sufficient condition for being non-Abelian has already been pointed out in the context of lattices \cite{Review-Goldman}, but other serious issues remain as we next discuss.

(V) The binary form of a commutator can lead to a natural but incorrect assumption that if a system evolves through two distinct closed loops A and B in parameter space with a common starting point, then if their order is reversed, the net value of the Wilson loop is unaffected for an Abelian gauge field, but generally differs for a non-Abelian one. However, the trace of a product of matrices is unchanged by a cyclic permutation so that ${\rm tr}[MN]={\rm tr}[NM]$, even when $MN\neq NM$. This means that two closed loops are \emph{not} sufficient to distinguish the non-Abelian nature via Wilson loop. At least three distinct loops are required. Numerical propagation in our model for $A_\theta, A_\alpha \neq 0$ demonstrates this in Fig.~\ref{Figure-4}. First, we compare the Wilson loops after two cycles, in forward B-C and reverse C-B sequence and find they agree in magnitude and phase at completion, though they can differ during evolution. Then, we compare the Wilson loops after three cycles, in sequence A-B-C with non-cyclic permutation A-C-B, there is a clear difference in the magnitude between the sequences at completion.

(VI) The phase of the Wilson loops is unaffected by the order of the loops, regardless of their number, as seen in Fig.~\ref{Figure-4}. That is because the Wilson matrix factorizes $W_\circ=W_\circ^{U(1)}W_\circ^{SU(2)}$, into $U(1)$ and $SU(2)$ components (see Eq~(\ref{Wilson-matrix}) below). The $SU(2)$ contribution to ${\cal W}$ is real, so the phase arises only from the Abelian $U(1)$ factor.

(VII) In the case of $U(2)$, the case most relevant for cold atom experiments, there is an important additional factor that has been overlooked. Consider the full Wilson matrix $W_\circ$  evaluated over two loops $A$ and $B$ with a common starting point, but the \emph{trace is not taken} and the argument in (II) above does not apply. It was stated in Ref.~\cite{Review-Goldman}, that $[W_\circ(A),W_\circ(B)]\neq 0$ would be a gauge invariant signature of genuinely non-Abelian structures. However, that does not always help for $N=2$. Factorization as mentioned in (III) implies the non-commutativity arises only from the $SU(2)$ part, but $SU(2)$ matrices are completely determined by their trace up to a unitary transformation (Appendix A). So, a gauge transformation $U$, which is a unitary transformation, can be found (Appendix A) such that $UW_\circ(A)W_\circ(B)U^\dagger=W_\circ(B)W_\circ(A)$. Therefore, when the trace is taken over the relevant density matrix,
\bn{\rm tr}\{\rho W_\circ(B)W_\circ(A)\}={\rm tr}\{U^\dagger \rho UW_\circ(A)W_\circ(B)\},\en and since for mixed states, density matrices satisfy $\rho\equiv U^\dagger \rho U$, in general even if $[W_\circ(A),W_\circ(B)]\neq 0$, we could still have ${\rm tr}\{\rho[W_\circ(A),W_\circ(B)]\}=0$. Three Wilson loops would be essential.

(VIII) The Wilson loop ${\cal W}$ is the invariant sum of the eigenvalues of $W_\circ$. The corresponding eigenvectors are the linear combination of the initial basis states that are unaltered by the path traversal apart from multiplication by the eigenvalues. The sum of the eigenvalues, hence the Wilson loop, thus measures the resulting `distortion'.

\begin{figure}[t]
\centering
\includegraphics[width=\columnwidth]{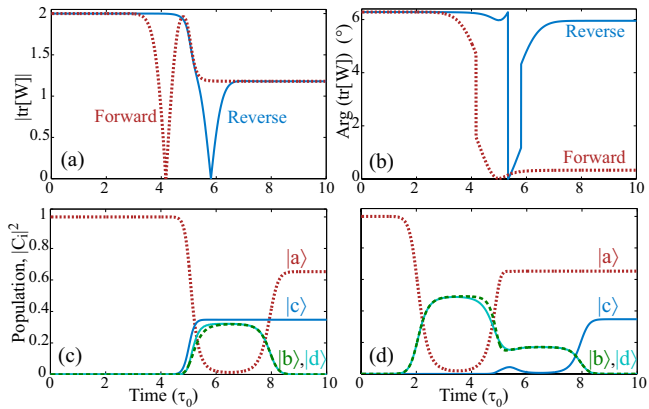}
\caption{(Color online) The Wilson integral over the loop B in Fig.~1, is plotted for forward (red dotted line) and reverse (solid blue line) evolutions: At end of the loop (a) the magnitudes are equal  but, (b) the phases are opposite ($2\pi-\theta\equiv -\theta$). The populations in the bare states are compared for (c) forward  and (d) reverse evolutions, showing that their final values match. All the population starts in state $|a\rangle$ (red dotted line) and is partially transferred to state $|c\rangle$ (solid blue line) at the end. The states $|b\rangle$ (green dashed line) and  $|d\rangle$  (solid cyan line) have almost identical evolutions, both starting and ending with vanishing occupation. }
\label{Figure-5}
\vspace{-.5cm}
\end{figure}

\section{Measuring the Wilson Loop Matrix} We will now show how the entire Wilson loop matrix, $W_\circ$ can be determined from the amplitudes of the bare states in our model. As an $U(2)$ matrix the Wilson loop can be parameterized as
\bn W_\circ={\begin{pmatrix}w_{11} &w_{12} \\w_{21} & w_{22} \\\end{pmatrix}}=e^{i\vartheta }{\begin{pmatrix}e^{i\vartheta _{1}}\cos \theta &e^{i\vartheta _{2}}\sin \theta \\-e^{-i\vartheta _{2}}\sin \theta & e^{-i\vartheta _{1}}\cos \theta \\\end{pmatrix}}.\label{Wilson-matrix}\en
For unique correspondence, the initial state is chosen to coincide with one of the dark states. At $t=0$ only $\delta\neq 0$, and since $q$ precedes both $p$ and $\alpha$, Eqs.~(\ref{parametrization}) and (\ref{non-Abelian-dark}) show that the initial state $C_i(0)=\delta_{a,i}$ coincides with $\Phi_1(0) $, likewise since $p$ is the last to vanish $\Phi_1(T)=-|c\rangle$ and $\Phi_2(T)=|a\rangle$.  Hence the amplitudes of those two bare levels yield the Wilson loop matrix elements $w_{11}$ and $w_{21}$.

However, that is insufficient for computing $W_\circ$ or even its trace which requires \emph{both} diagonal elements: While the sequence measures $\vartheta\pm \vartheta_{1(2)}$, without at least one more independent measurement we cannot determine $\vartheta\mp \vartheta_{1(2)}$. But, our choice of time evolution was deliberate to provide a solution:  We simply run the cycle backwards with the same initial condition $C_i(0)=\delta_{a,i}$, in which case, at $t=0$, $\delta\neq 0$ as before, but now $p$ precedes both $q$ and $\alpha$, so the initial state coincides with $\Phi_2(0) $, and since $q$ is the last to vanish, $\Phi_1(T) =|a\rangle$ and $\Phi_2(T) =|c\rangle$. Hence the amplitudes of those two bare levels yield the matrix elements $w'_{12}$ and $w'_{22}$ of  the inverse matrix $W_\circ^{-1}=W_\circ^\dagger$, which are related to the original matrix elements as $w'_{12}=w_{21}^*$ and $w'_{22}=w_{22}^*$.  This determines $w_{22}$ and $w_{12}=-w_{21}^*\exp(i{\rm Arg}[w_{11}w_{22}])$, and hence the complete matrix $W_\circ$ and its trace the Wilson loop ${\cal W}$.  Simulations for loop B in Fig.~\ref{Figure-2} confirm this
\scriptsize
\bn  \Psi^F(T)&=& ({\color{blue} 0.626 + 0.510i},  0,  {\color{red} -0.570 - 0.152i},  0)
\n\\
W_\circ^F(T)&=& \left(
\begin{array}{cc}
 {\color{red} 0.570 + 0.155i} & -0.806 + 0.025i  \\
 {\color{blue} 0.622 + 0.513i}  & 0.546 + 0.227i
\end{array}\right)\n\\
\Psi^R(T)&=&  ({\color{blue} 0.626 - 0.510i},0,   {\color{red} 0.544 - 0.226i},0)
\n\\
W_\circ^R(T)&=& \left(
\begin{array}{cc}
 0.570 - 0.155i &  {\color{blue} 0.622 - 0.513i} \\
 -0.806 - 0.025i  & {\color{red} 0.546 - 0.227i}
\end{array}
\right).\en
\normalsize
Here $\Psi(T)$ are obtained by evolution with the Hamiltonian in Eq.~(\ref{Hamiltonian}) forward (F) and in reverse (R) and the Wilson matrices are computed with adiabatic equations Eq.~(\ref{evolution-equations}). The agreement of the specific elements are evident exactly as discussed above. The time evolution of ${\cal W}$ in both forward and reverse is plotted in Fig.~\ref{Figure-5} and at the end of the cycles the magnitudes match, while the phases are complex conjugates as expected. The bare level populations are also plotted and are in agreement.

\section{Physical Realizations} While much of our discussion of Wilson loops is general in scope, we now discuss methods for implementation of the Hamiltonian in Eq.~(\ref{Hamiltonian}) in experiments to test our results with cold atoms. Our model is distinct from the popular multipod schemes  \cite{Bergmann-PRA-STIRAP-geometric, Fleischhauer-PRL-non-abelian-gauge,Clark-zitterbewegung,Juzeliunas-double-refraction,Galitski-spin-orbit} and the ring-coupling scheme favored in recent experiments \cite{Spielman-magnetic, Spielman-Zitterbewegung, Spielman-non-Abelian-monopole}, and is particularly suited for examining group structure as it allows continuous variation form Abelian to non-Abelian structures within one configuration.

The closed loop configuration imposes non-trivial constraints which can be met by our scheme, as we now show in the context of possible experiments. Consider four electronic levels of an atom labeled $\alpha\in\{a,b,c,d\}$, with bare eigenstates $H_0\psi=\hbar\omega_\alpha |\alpha\rangle$, which are coupled by electromagnetic fields of frequencies $\nu_i, i\in\{1,2,3,4\}$ in the sequence $a\leftrightarrow b\leftrightarrow c\leftrightarrow d\leftrightarrow a$. Expressing the state vector in this basis
\bn |\psi(t) \rangle=\sum_\alpha c_\alpha(t) e^{-i\omega_\alpha t} |\alpha\rangle \en
we insert into the Schr\"odinger equation including the field potentials, $i\hbar\partial_t|\psi\rangle=(H_0+V)|\psi\rangle$. Using a rotating wave approximation, and transforming to rotating frame $c_i(t)=C_i(t) e^{-i\phi_i(t)}$ eliminates the exponentials and yields four coupled equations  $i\hbar\partial_t C_i=H_{ij}C_j$.

\begin{figure}[t]
\centering
\includegraphics[width=\columnwidth]{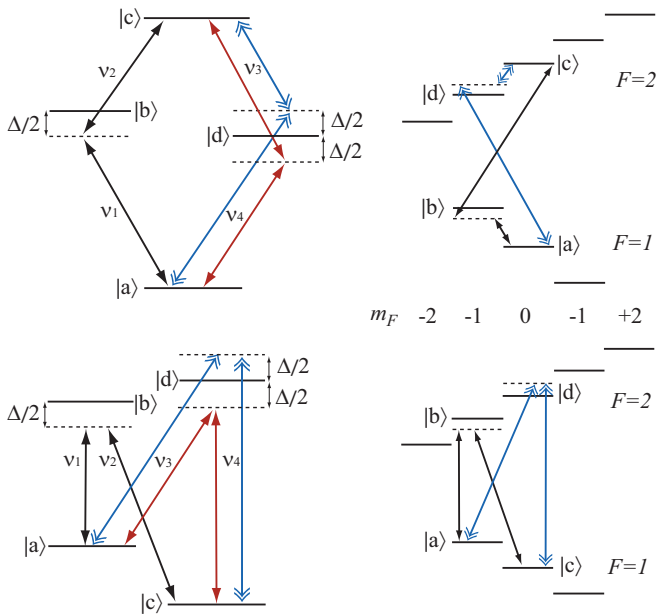}
\caption{(Color online) Two possible implementations, with \emph{upper} figures corresponding to $H_1$ in Eq.~(\ref{phyisical-hamiltonians1}) and \emph{lower} to $H_2$ in Eq.~(\ref{phyisical-hamiltonians2}).  \emph{Left}: In each level scheme, the non-degenerate ($H_{44}=+\hbar\Delta/2$) case is shown with single-headed red arrows, and the degenerate case ($H_{44}=-\hbar\Delta/2$) with double-headed blue arrows. \emph{Right}: The corresponding couplings for hyperfine states of alkali atoms (only degenerate case shown).}
\label{Figure-1}
\end{figure}

The effective $4\times4$ Hamiltonian is defined by the complex Rabi frequencies
$\hbar\Omega_{i}=-{\cal E}_i e\langle \beta|x|\alpha\rangle$ and the detunings $\Delta_i=(\omega_{\beta}-\omega_\alpha-\nu_i) $, where $\nu_i$ is the appropriate coupling field for $\alpha \leftrightarrow \beta$ and $\omega_\beta>\omega_\alpha$ is assumed. There are multiple possible choices to arrive at the Hamiltonian in Eq.~(\ref{Hamiltonian}), here we present two examples for both of which the closed loop requires $(\nu_1+\nu_2)=(\nu_3+\nu_4)$.

For level ordering $\omega_a<\omega_b<\omega_c$ and $\omega_a<\omega_d<\omega_c$, the detunings need to satisfy  $\Delta_1+\Delta_2-\Delta_3-\Delta_4=0$ and we get a diamond configuration, $H_1$, corresponding to the choice $\Delta_1=-\Delta_2=\pm\Delta_3=\Delta/2$ (upper sign for non-Abelian).

\bn H_1={\textstyle\frac{\hbar}{2}}\begin{bmatrix} 0 & \Omega_1^*& 0&\Omega_4^* \\ \Omega_1 & \Delta& \Omega_2^*&0\\
0 & \Omega_2& 0&\Omega_3 \\ \Omega_4 & 0& \Omega_3^*&\mp\Delta\end{bmatrix}.\label{phyisical-hamiltonians1}\en

If, instead, we choose $\omega_c<\omega_a<\omega_b<\omega_d$, the detunings need to satisfy $\Delta_1-\Delta_2+\Delta_3-\Delta_4=0$ and we get a folded diamond  configuration, $H_2$, with the choice of $\Delta_1=\Delta_2=\mp \Delta_3=\Delta/2$, we get:

\bn
H_2={\textstyle\frac{\hbar}{2}} \begin{bmatrix} 0 & \Omega_1^*& 0&\Omega_4^* \\ \Omega_1 & \Delta& \Omega_2 &0\\
0 & \Omega_2^*& 0&\Omega_3^* \\ \Omega_4 & 0& \Omega_3&\mp\Delta\end{bmatrix}.\label{phyisical-hamiltonians2}\en

Both options could be implemented within the $F=1$ and $F=2$ hyperfine levels of alkali atoms such as $^{87}Rb, ^{39}K, ^{41}K$, as illustrated in Fig.~\ref{Figure-1} along with the level diagrams. Of the two, $H_2$ may be easier to implement, using all microwave fields in relatively low magnetic field. In order to implement $H_1$, besides two microwave fields, radio-frequency would be required for the intra-level transitions, and as such sufficient Zeeman splitting would call for large magnetic fields. For either case, the Rabi frequencies will need to be precisely controlled, their magnitudes need to satisfy $|\Omega_1| =|\Omega_4|$  and  $|\Omega_2| =|\Omega_3|$ which could be achieved by controlling the field intensities, and  their relative phases need to sum to zero. Each coupling could also alternately be achieved with Raman transitions of a pair of laser fields.

\section{Conclusions} We described how a Wilson loop can be used to distinguish non-Abelian gauge structures, specially noting the necessity for three distinct loops, and for U(2), even with the full Wilson loop matrix. We provided a method to measure the full matrix without requiring a lattice. For that purpose, we introduced a novel cyclic model which can have broader utility as it allows realizing both $U(1)$ and $U(2)$ gauge phenomena in a single configuration, and we showed viable implementation in the ground state manifold of alkali atoms.

\section{Acknowledgments} We acknowledge discussions with S. Aubin and M. Gajdacz, the support of NSF under Grants No. PHY-1313871 and PHY11-25915, and the KITP at UCSB where this work began under a Kavli Scholarship.

\appendix

\section{Lemmas for U(2) Wilson Loop Matrix}

Here we prove a pair of Lemmas used in item (VII) in Sec.~VII.\\

{\bf Lemma 1}: \emph{Two SU(2) matrices with the same trace, are unitarily  equivalent.}\vs{2mm}

{\bf Proof}: Let the two matrices be $M$ and $N$, their characteristic equations are identical
\bn \lambda^2-{\rm tr}[M]+1=0\en
since for $SU(2)$ the determinant is 1, and ${\rm tr}[M]={\rm tr}[N]$. It follows therefore, that they have the same eigenvalues. Being unitary, they both have diagonal representations
\bn  M=U_1DU_1^\dagger \h{2cm}N=U_2DU_2^\dagger\en
where $D$ can be identical for both, with eigenvalues along the diagonals, and $U_1$ and $U_2$ are unitary matrices formed from the eigenvectors. Therefore,
\bn  U_1^\dagger MU_1&=&U_2^\dagger NU_2=D\n\\
\Rightarrow UMU^\dagger&=&N, \h{1cm}{\rm where}\ U=U_2U_1^\dagger\en
Hence, $M$ and $N$ unitarily equivalent, that is they are similar upto a unitary transformation.

\h{5mm}

{\bf Lemma 2}:  \emph{For two U(2) Wilson loop matrices, $W_\circ(A)$ and $W_\circ(B)$, following two separate loops A and B, starting and ending at the same point in parameter space, the products of the matrices in mutually reversed orders $W_\circ(A)W_\circ(B)$ and $W_\circ(B)W_\circ(A)$ are unitarily  equivalent.}
\vs{2mm}

{\bf Proof}:
Assume the two products are not equal
\bn  W_\circ(A)W_\circ(B)\neq W_\circ(B)W_\circ(A).\en
Then since $W_\circ(A)$ and $W_\circ(A)$ are unitary, they can be be factorized into $U(1)$ and $SU(2)$ components
\bn W_\circ(A)=W_\circ^{(U)}(A)W_\circ^{(SU)}(A).\en
Since the $U(1)$ part is just a phase factor, those factors would commute
\bn W_\circ^{(U)}(A)W_\circ^{(U)}(B)= W_\circ^{(U)}(B)W_\circ^{(U)}(A).\en
Therefore the inequality above must arise only from the $SU(2)$ factors
\bn  W_\circ^{(SU)}(A)W_\circ^{(SU)}(B)\neq W_\circ^{(SU)}(B)W_\circ^{(SU)}(A).\en
However, since a binary product of matrices have the same trace regardless of permutation
\bn  {\rm tr}[W_\circ^{(SU)}(A)W_\circ^{(SU)}(B)]={\rm tr}[W_\circ^{(SU)}(B)W_\circ^{(SU)}(A)]\en
and the product of two $SU(2)$ matrices is another SU(2) matrix by their group closure property, therefore by \emph{Lemma 1}, it follows that some unitary matrix $U$ can be found such that
\bn  UW_\circ^{(SU)}(A)W_\circ^{(SU)}(B)U^\dagger=W_\circ^{(SU)}(B)W_\circ^{(SU)}(A).\en
Then, multiplication by the $U(1)$ factors on both sides leads to
\bn  UW_\circ(A)W_\circ(B)U^\dagger=W_\circ(B)W_\circ(A).\en

\end{document}